\documentstyle[preprint,aps,epsfig]{revtex}
\newcommand{\lsim}{\mathrel{\mathop{\kern 0pt \rlap
  {\raise.2ex\hbox{$<$}}}
  \lower.9ex\hbox{\kern-.190em $\sim$}}}
\newcommand{\gsim}{\mathrel{\mathop{\kern 0pt \rlap
  {\raise.2ex\hbox{$>$}}}
  \lower.9ex\hbox{\kern-.190em $\sim$}}}
\newcommand{\R}     {{\mathcal R}}

\newcommand{\be}     {\begin{equation}}
\newcommand{\ee}     {\end{equation}}
\newcommand{\bea}     {\begin{eqnarray}}
\newcommand{\eea}     {\end{eqnarray}}
\newcommand{\no}     {\nonumber}
\newcommand{\Lm}     {\Lambda}
\newcommand{\Gm}     {\Gamma}

\begin{document}
\draft
\preprint{
\vbox{\hbox{\bf (hep-ph/0009245)}
      \hbox{SNUTP\hspace*{.2em}00-026}}
}
\title{
Radion effects on the production  \\
of an intermediate-mass scalar and $Z$ at LEP II
}

\author{
Seong Chan Park,~~ H.~S.~ Song and~~ Jeonghyeon~
Song }
\vspace{1.5cm}
\address{
Department of Physics, Seoul National University,
Seoul 151-742, Korea
}

\maketitle
\thispagestyle{empty}
\setcounter{page}{1}

\begin{abstract}
\noindent 
We have studied the $ e^+ e^- \to Z \phi_i \to Z jj$ process,
where $\phi_i$ is the Higgs and/or radion bosons.
The implications of the radion effects
on the preliminary ALEPH data are also discussed.
The case of the lighter radion than Higgs boson
is disfavored by the ALEPH analyses of the $b$ tagged four-jet data,
since the radion predominantly decays into two gluon jets 
due to the QCD trace anomaly.
If the radion is highly degenerate in mass
with the Higgs,
the cross section can be increased
more than at one sigma level,
with natural scale of the vacuum expectation value
of the radion.
\end{abstract}

\newpage

\section{Introduction}

The standard model (SM) has been 
very successful in describing the known
electroweak interactions
of gauge bosons and fermions.
Nevertheless one of the most crucial ingredients of the SM, 
the Higgs boson, has not been experimentally discovered yet\cite{Higgs}.
The Higgs mechanism is responsible for the electroweak
symmetry breaking in the SM,
of which the effects should appear
by the scale below $(8\pi \sqrt{2}/3 G_F)^{1/2}\sim 1$ TeV
for the preservation of unitarity
in the $W_L W_L \to W_L W_L$ process\cite{Duncan:1986vj}.
Hence the primary
efforts of the future collider experiments are to be directed
toward the search for Higgs bosons.

The search techniques for Higgs bosons according to
the Higgs mass range have been extensively studied in the literature
\cite{Gunion:1988ke}.
Light Higgs bosons with mass below 107.7 GeV
are excluded by LEP-II experiments at the 95\%
confidence level\cite{HiggsLow}.
For $heavy$ Higgs bosons $(m_h \gsim 2 m_Z)$,
the experimental search is rather straightforward
by observing the Higgs decay into a $Z$ boson pair,
or by studying $W W$ scattering at $e^+ e^-$ 
or hadron colliders\cite{heavy}.
For the $intermediate$-mass $( m_Z< m_h \lsim 2 m_Z)$,
one of the best reactions for their detection 
is known to be
$ e^+ e^- \to Z^* \to Z h$ (see the corresponding
Feynman diagram in Fig.~1)\cite{light}.
The Higgs boson in this mass range
decays mainly into $b \bar{b}$.
The direct physics background from the continuum production of
$ e^+ e^- \to  Z b \bar{b} $
has been also discussed in the literature.

Theoretically,
the Higgs boson holds a distinctive position
in understanding the physics beyond the SM.
The existence of the Higgs boson in the SM,
as a fundamental scalar particle, causes
the well-known gauge hierarchy problem:
It is unnatural that the
Higgs mass at the electroweak scale is protected from the
presence of the enormous Planck scale.
This gauge hierarchy problem is the main motivation
for various models of new physics,
such as supersymmetric models, technicolor models, and extra dimensional 
models.
Among these,
an extra-dimensional model recently
proposed by Randall and Sundrum (RS)
requires the existence of another scalar field,
called the radion,
of which the mass can be compatible with the Higgs mass\cite{RS}.

In this report,
we study the radion effects on the
$ e^+ e^- \to Z \phi_i \to Z jj$
process,
where the $\phi_i$ is a Higgs boson and/or radion,
and $j$ denotes a hadronic jet.
Recently, an exciting, even preliminary,
news has come from the ALEPH group\cite{ALEPH}.
From the analyses of the 237 pb$^{-1}$ of data collected
at $\sqrt{s}$ up to 209 GeV,
the ALEPH has measured 
three standard deviation from the
SM continuum background.
This result is compatible with the SM Higgs of mass
around 114 GeV.
Another unusual result is that
the cross section with the SM Higgs is still
high; the standard deviation is 
between $1 \sigma$ to $2.5 \sigma$\cite{ALEPH}.
If this result remains valid in the future,
the radion,
if degenerate in mass with the Higgs, 
can be one of the best candidates 
for the explanation of the excess.
We caution the reader that
due to the small number of events and the presence
of background in the current ALEPH data,
it is very impetuous to draw any physical conclusion
at this moment.

In the RS model,
the hierarchy problem is solved by
a geometrical exponential factor, called a warp factor \cite{RS}.
The spacetime has a single $S^1/Z_2$
orbifold extra dimension with the metric
\begin{equation}
d s^2= e^{-2 k r_c |\varphi| } \eta_{\mu\nu} d x^\mu d x^\nu
+ r_c^2 d \varphi^2,
\end{equation}
where the $\varphi$ is confined to $0 \leq |\varphi| \leq \pi$.
Two orbifold fixed points accommodate two three-branes,
the hidden brane at $\varphi=0$ and
our visible brane at $|\varphi|=\pi$.
The allocation of our brane at $|\varphi|=\pi$
renders a fundamental scale $m_0$ to appear as
the four-dimensional physical mass $m=e^{-k r_c \pi} m_0$.
The hierarchy problem can be answered
if $k r_c \simeq 12$.
In this model,
very critical is the stabilization of
the compactification radius $r_c$:
Without a stabilized radius,
we should impose a fine-tuning constraint
between the matter densities
on the two branes,
which causes non-conventional cosmologies\cite{R-cos}.
Several scenarios for the stabilization mechanism
have been proposed,
where the radion modulus $\phi$ 
can be significantly lighter than the 
geometrically suppressed Planck scale on the visible brane, 
$\Lm \sim {\mathcal O} ( {\rm TeV}) $ \cite{gw}.
Various phenomenological aspects of the radion at colliders
have been studied in the literature:
The decay modes of the radion according to its mass range
are different from those of the Higgs
(e.g., its dominant decay mode with
mass below $2 m_Z$ is into two gluons) \cite{Ko};
without a curvature-scalar Higgs mixing,
the radion effects
on the oblique parameters
of the electroweak precision observations are small\cite{Csaki:2000zn};
the electroweak mixing angle can undergo a substantial
correction if the mass and vacuum expectation value (VEV)
of the radion are below TeV\cite{Kim:2000ks};
the radion effects on the phenomenology at low energy
colliders\cite{Mahanta:2000zp}
and at high energy colliders\cite{Mahanta:2000ci}
have been also discussed
(e.g., a lower bound of 35 GeV on the radion mass
is obtained from the LEP-I bound on the 60 GeV Higgs mass).

The interactions of the radion  $\phi$
with the SM particles show
similar behaviors to those of the Higgs boson.
The interaction Lagrangian is
\begin{equation}
{\mathcal L}
= \frac{1}{\Lm_\phi} ~\phi~ T^\mu_\mu
\,,
\end{equation}
where the $T^\mu_\mu$ is the trace of the conserved 
and symmetrized energy-momentum tensor of the SM fields,
and the $\Lm_\phi$ is the VEV
of the radion.
The coupling of the radion with a fermion or
gauge boson pair
is the same as that of the Higgs,
except for a factor of $(v/\Lm_\phi)$.
The $v$ is the VEV of the Higgs.
The massless gluons and photons also contribute
to the $T^\mu_\mu$,
due to the trace anomaly.
It is known that the trace anomaly appears
since the scale invariance of massless fields
is broken by the running of gauge couplings\cite{Collins:1977yq}.
Thus the interaction Lagrangian between two gluons and
the Higgs or the radion is
\begin{equation}
{\mathcal L}_{h(\phi)-g-g}
=
\left[
\left( \frac{v}{\Lm_\phi} \right)
\left\{
b_3 + I_{1/2}(z_t)
\right\} \phi
+
I_{1/2}(z_t) h 
\right]
\frac{\alpha_s}{8 \pi v} \,{\rm Tr}
(G^C_{\mu\nu}G^{C \mu\nu} )
\,,
\end{equation}
where $z_t = 4 m_t^2/m_h^2$,
and
the QCD beta function coefficient is
$b_3 =11-2 n_f /3$ with the number of dynamical quarks $n_f$.
The loop function $ I_{1/2}(z) $ is defined by
\be
I_{1/2}(z) =z [ 1+ (1-z) f(z)]
\,,
\ee
where the $f(z)$ is
\begin{eqnarray} 
\label{ffnt}
f(z) 
& = & \left\{  \begin{array}{cl}
           {\rm \arcsin}^2(1/\sqrt{z}) \,,   &\qquad z \geq 1\,, \\
           -\frac{1}{4}\left[\ln \left(\frac{1+\sqrt{1-z}}{
                                             1-\sqrt{1-z}}\right)
                             -i\pi\right]^2\,, & \qquad z\leq 1\,.
               \end{array}
\right.
\end{eqnarray}
It is to be noted that
the phenomenology of radions 
can be determined by two parameters,
$m_\phi$ and $\Lm_\phi$.
For the intermediate-mass radion,
the dominant decay rate into $gg$ is larger than that into 
$b\bar{b}$ by an order of magnitude.

Let us now review the ALEPH results\cite{ALEPH}.
By using both a neural-network-based stream (NN) and
a cut-based stream,
the ALEPH group analyzed the final states most
from the process $e^+ e^- \to H Z$:
the four-jet final states ($H q\bar{q}$ ),
the missing energy final states ($H\nu\bar{\nu}$ ),
the lepton pair final states
($H l^+ l^-$ where $l$ denotes an electron or muon),
and the tau lepton final states
($H\tau^+ \tau^-$ and $H\to \tau^+ \tau^-,\,Z\to q\bar{q}$).
Common discriminant for the event selection is the
reconstructed Higgs boson mass.
As a second discriminant,
the neural network output is used for the four-jet NN analysis
while the sum of the $b$ tagging neural network output
values is used for the missing energy NN and lepton pair selections.

We are interested in the radion effects on the circumstance
where the lighter scalar is produced on-shell.
If the radion is much heavier than the Higgs boson,
the ordinary analyses with the SM Higgs remain intact.
The opposite possibility of the presence of lighter radions
is disfavored,
since the ALEPH analysis of the $b$-tagged data is compatible
with the SM Higgs boson hypothesis,
while a radion with mass around 114 GeV predominantly decays 
into two gluons.

If the radion is almost degenerate
in mass compared with the Higgs,
the cross section for the process 
$e^+ e^- \to Z \phi_i \to Z jj$ is to be modified.
With the degenerate radion involved,
two hadron jets can be either $b\bar{b}$ jets or $g g$ jets,
while the SM Higgs boson at  114 GeV mass predominantly decay into 
$b\bar{b}$ jets.
The radion effect is casted into the ratio between the total cross
sections with and without the radion involved:
\bea
\label{R}
{{\sigma^{Zjj} _{{ h}+\phi}} \over{\sigma ^{Zjj}_{ h}}}
  &\simeq & 
\frac{
\sigma^{Zb\bar{b}} _{ h+\phi}
+ \sigma^{Z g g} _{ \phi} 
} { 
\sigma^{Zb\bar{b}}_h 
}
\\ \no
&=&
\frac{ 
\left| M_{e^+ e^- \to Zh} D_h M _{h\to b\bar{b}}
+ M_{e^+ e^- \to Z \phi } D_\phi M _{\phi \to b\bar{b}}
\right|^2
}{
\left| M_{e^+ e^- \to Zh} D_h M _{h\to b\bar{b}}\right|^2
}
+ \frac{
\sigma( e^+ e^- \to Z \phi ) Br ( \phi \to gg)
}{
\sigma( e^+ e^- \to Z h ) Br ( h \to b\bar{b} )
}
\\ \no
&=& 1 + 2 \left({v \over {\Lambda_\phi}}\right)^2
     {\Re}e \left( {{D_\phi} \over {D_h}} \right)
+ \left({v \over {\Lambda_\phi}}\right)^4
\left|
{{D_\phi} \over {D_h}}
\right|^2
+ \left({v \over {\Lambda_\phi}}\right)^2
\frac{ \Gamma_h}{\Gamma_\phi}
\frac{ \Gamma_{\phi \to g g} }{\Gamma_{h \to b\bar{b} } }
\,,
\eea
where $M$ is the scattering amplitude,
and $D_i ( q^2 )$ is the propagation factor for the
$i$-th scalar particle.
Since the narrow width approximation,
$|D_i|^2 \simeq ( \pi / \Gamma_i m_i) \delta(s-m_i^2)$,
 with the degenerate assumption
$(m_h \approx m_\phi)$ implies
\be
{\Re}e \left( {{D_\phi} \over {D_h}} \right)
=
\left.
\frac{
(s-m_h^2)(s-m_\phi^2) + \Gm_h \Gm_\phi m_h m_\phi
}
{
(s-m_\phi^2)^2 + \Gm_\phi^2  m_\phi^2
}
\right|_{(\sqrt{s}\, \approx \,m_h \approx\, m_\phi)}
\approx
\sqrt{
\frac{\Gm_h}{\Gm_\phi}
}
\,,
\ee
the ratio become
\bea
\label{Zjj}
{{\sigma^{Zjj} _{{ h}+\phi}} \over{\sigma ^{Zjj}_{ h}}}
&=& 1 + 2 \left({v \over {\Lambda_\phi}}\right)\sqrt{\R}
+ \left({v \over {\Lambda_\phi}}\right)^2 \R
\left[
1+\frac{\alpha_s}{12\pi^2} \frac{m_h^2}{m_b^2}
\frac{|b_3+I_{1/2}(z_t)|^2}{(1-z_b)^{3/2} }
\right]
\,.
\eea
From the fact that the decay-rate ratio
${\Gm_h}/{\Gm_\phi}$ is inversely proportional
to $(v/\Lm_\phi)^2$, we have defined $\R$ in Eq.~(\ref{Zjj}) by
\be
\R \equiv
\left(
\frac{v}{\Lm_\phi}
\right)^2
\frac{\Gamma_h}{\Gamma_\phi}
\simeq
\frac{ 1+ C_{\tau b} + C_{\rm QCD}(M_{j j}) | I_{1/2}(z_t ) |^2}
{ 1+ C_{\tau b} + C_{\rm QCD}(M_{j j}) |b_3+ I_{1/2}(z_t ) |^2}
\,,
\ee
where $M_{j j}$ is the reconstructed scalar mass,
$z_x=4 m_x^2/M_{j j}^2$, and
the $C_{\tau b}(M_{j j})$ and $ C_{\rm QCD}(M_{j j})$ are,
\bea
C_{\tau b} &=& \frac{1}{3}
\left(
\frac{m_\tau}{m_b}
\right)^2
\left(
\frac{1-z_\tau}{1-z_b}
\right)^{3/2}
\,,
\\ \nonumber
C_{\rm QCD}(M_{j j})
&=&
\frac{\alpha_s^2}{12\pi^2}
\left(
\frac{M_{j j}}{m_b}
\right)^2
\,.
\eea
Here we have taken into account of
$b\bar{b}$, $\tau^+\tau^-$,
and $g g$ as the decay modes of an
intermediate-mass scalar particle.
The difference between the Higgs and radion masses
is assumed to be below
the mass resolution at $e^+ e^-$ colliders.
Since the QCD beta function coefficient $b_3$ is 7
for $n_f=6$,
the $\R$
is less than one.
Thus the ratio
$ \sigma^{Zjj} _{h+\phi}/ \sigma^{Zjj} _{h}$
is larger than one,
implying that
the presence of degenerate radions 
increases the cross section.
And the enhancement can be substantial since $\R (m_h =114 ~
{\rm GeV} )=0.08$.

Now we discuss the implications of the radion effects
for the ALEPH results.
In this degenerate case,
it is possible to explain one of the preliminary
ALEPH results;
the cross section is measured still higher
than the SM Higgs prediction. 
The $1\sigma$ and
$2\sigma$ excesses in the cross section at $\sqrt{s}=207$ GeV
and $\int {\mathcal L} dt = 216.1$ pb$^{-1}$
can be explained by Higgs-radion
degeneracy
with $\Lm_\phi=2.2$ TeV and $\Lm_\phi=1.4$ TeV,
respectively.
The Higgs mass $m_h$ is set to be 114 GeV,
used as the QCD scale in $\alpha_s$.
When the $b$ tagging of the jets
from the Higgs decay is highly performed
as in the high purity data selection of the ALEPH\cite{ALEPH},
the relevant quantity to signal the radion effects
is 
\be
\label{Rb}
{{\sigma ^{Zb\bar{b}}_{{h}+\phi}} 
\over{\sigma^{Zb\bar{b}} _{h}}}
=
 1 + 2 \left({v \over {\Lambda_\phi}}\right)\sqrt{\R}
+ \left({v \over {\Lambda_\phi}}\right)^2 \R
\,.
\ee
For $\Lm_\phi=2.2$ TeV which would yield 1$\sigma$
excess in the 216.1 pb$^{-1}$ data,
the radion effects would lead to 0.34 $\sigma$ excess
in the cross section compared to that 
without the radion effects.
When the luminosity increases into 700 pb$^{-1}$,
the $1\sigma$ and
$2\sigma$ excesses in the cross section correspond to
$\Lm_\phi=3.1$ TeV and $\Lm_\phi=2.0$ TeV, respectively.
It is concluded that highly degenerate radions
with the SM Higgs can be a good explanation
for the excess of the scalar production
accompanying $Z$ at $e^+ e^-$ colliders.

In summary,
we studied the
$ e^+ e^- \to Z \phi_i \to Z jj$
process,
where $\phi_i$ is the Higgs and/or radion particles.
In the case of the lighter radion than the Higgs,
the QCD trace anomaly renders the radion to decay dominantly into
$gg$,
which is not compatible with the ALEPH $b$-tagged results.
If the radion is highly degenerate in mass
with the Higgs,
the cross section can be increased
more than at one sigma level.
This can be one of the most natural explanations
for the preliminary ALEPH results
of the excess in the cross section
with the SM Higgs.

\acknowledgments

\noindent
The work was supported by the BK21 Program.

\def\IJMP #1 #2 #3 {Int. J. Mod. Phys. A {\bf#1},\ #2 (#3)}
\def\MPL #1 #2 #3 {Mod. Phys. Lett. A {\bf#1},\ #2 (#3)}
\def\NPB #1 #2 #3 {Nucl. Phys. {\bf#1},\ #2 (#3)}
\def\PLBold #1 #2 #3 {Phys. Lett. {\bf#1},\ #2 (#3)}
\def\PLB #1 #2 #3 {Phys. Lett.  {\bf#1},\ #2 (#3)}
\def\PR #1 #2 #3 {Phys. Rep. {\bf#1},\ #2 (#3)}
\def\PRD #1 #2 #3 {Phys. Rev. {\bf D}{\bf#1},\ #2 (#3)}
\def\PRL #1 #2 #3 {Phys. Rev. Lett. {\bf#1},\ #2 (#3)}
\def\PTT #1 #2 #3 {Prog. Theor. Phys. {\bf#1},\ #2 (#3)}
\def\RMP #1 #2 #3 {Rev. Mod. Phys. {\bf#1},\ #2 (#3)}
\def\ZPC #1 #2 #3 {Z. Phys. C {\bf#1},\ #2 (#3)}
\def\EPJ #1 #2 #3 {Euro. Phys. Jour. {\bf#1},\ #2 (#3)}

\begin{center}
\begin{figure}[htb]
\smallskip
\smallskip
\smallskip
\smallskip
\smallskip
\hbox to\textwidth{\hss\epsfig{file=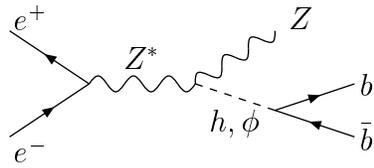}\hss}
\smallskip
\caption{ The Feynman diagram of the 
$ e^+ e^- \to Z \phi_i \to Z b\bar{b}$
process with the Higgs and radion.
}
\label{fig1}
\end{figure}
\end{center}

\end{document}